\documentclass[twocolumn,showpacs,preprintnumbers,amsmath,amssymb]{revtex4}

\usepackage{graphicx}
\usepackage{dcolumn}
\usepackage{bm}

\begin{document}

\preprint{}

\title{Bouncing Neutrons and the Neutron Centrifuge}

\author{P. J. S. Watson\\}
 \altaffiliation[Also at ]{Department of Theoretical Physics, University of Oxford}
 \email{watson@physics.carleton.ca}
\affiliation{Physics Dept, Carleton University, \\ Ottawa, Canada, K1S 5B6
}

\date{\today}

\begin{abstract}
The recent observation of the quantum state of the neutron bouncing freely under gravity allows some novel experiments. A method of purifying the ground state is given, and possible applications to the measurement of the electric dipole moment of the neutron and the short distance behaviour of gravity are discussed.
\end{abstract}

\maketitle

\section{\label{sec:level1}Introduction:}
In an elegant experiment, Nesvizhevsky et al \cite {Nes} \cite{Nes2} have observed the quantum states of a neutron bouncing freely under gravity. Their experiment shows that the state exists with approximately the predicted properties by observing its extinction by lowering a neutron absorber.  In this paper, we discuss a method for purifying the ground state and then discuss experiments to exploit it.

	Ultracold neutron (UCN) technology \cite{Golub} has been around for many years, and it is exploited in the latest EDM experiments \cite{Ramsey}. This relies on the observation that (for example) beryllium metal totally reflects neutrons at a temperature of below 100 K. This implies the existence of the quantum state; that of a neutron bouncing freely under gravity. Although the idea is a fairly obvious one, and the theory is very old \cite{Flugge}, \cite{Romo}, it appears that no attempts were made to observe it until recently. It seems possible that if the ground state of the system could be isolated, there are several experiments which could be performed

In the next section we show that a simple device based on a classical centrifuge could be used to purify the ground state. We then show that the lifetime of the state is expected to be long. We then consider two possible applications to fundamental physics. Firstly, it is well-known that the breakdown of CP invariance and hence T invariance in fundamental interactions implies the existence of an electric dipole moment (EDM) for the neutron. Predictions  of the magnitude vary from $10^{-25}$ e-cm down to $10^{-35}$ e-cm, while the current limit is $0.5 \times 10^{-25}$ e cm. Obviously it is of considerable interest to lower this, since it restricts possible extensions of the standard model considerably. Finally we investigate the possibility of using the system to test for extra gravitational interactions at very short range.

\section{The bouncing neutron}
	In this section we review the theory for the one-dimensional case. The bounce eigenfunctions $Z_n(z)$ are defined by the Schrodinger equation which describes the neutron under gravity is
\begin{equation}-{{\hbar ^2} \over {2m}}{{d^2} \over {dz^2}}Z\left( z \right)+\left( {\sigma z-E} \right)Z\left( z \right)=0\  ,\  \sigma =mg \end{equation}
This can be converted to a dimensionless form via the substitution
\begin{equation}y=\beta z-y_n\   ,\   \beta =\left( {{{2m\sigma } \over {\hbar ^2}}} \right)^{1/3} \end{equation}
The boundary condition for a totally reflecting "ground" is that the wave-function vanishes at z = 0. This gives
\begin{equation}{{d^2} \over {dy^2}}Z_n\left( y \right)+(y-y_n)Z_n\left( y \right)=0\end{equation}

with 
\begin{equation}Z_n\left( {-y_n} \right)=0\  \end{equation}
As is well known, this is the equation for the Airy function, and hence  $y_n$ is the n-th zero of the function. This gives rise to an energy $E_n$ and a "bounce height" $z_n$
\begin{equation}E_n={{\hbar ^2\beta ^2} \over {2m}}y_n\  ,\  z_n={{y_n} \over \beta }\end{equation}
For the neutron, the ground state parameters are $E_0=1.407 peV$
 and $\  z_0=13.7\mu $ (these results differ slightly from those quoted in \cite{Nes2}. The latter number implies that one could build a macroscopic apparatus sensitive to the parameters of a quantum state. 
	The usual method of looking for bound states, by measuring the spectrum of emitted photons, is not available. The frequency for the 1-photon transition between the first excited and the ground state is  254 Hz \footnote {It might be noted that this is almost exactly middle C. There is unlikely to be any significance in this observation} and one can estimate the radiative lifetime of the first excited state to be much larger than the Hubble time!

 \section{The Neutron Centrifuge}
\begin{figure}
\includegraphics{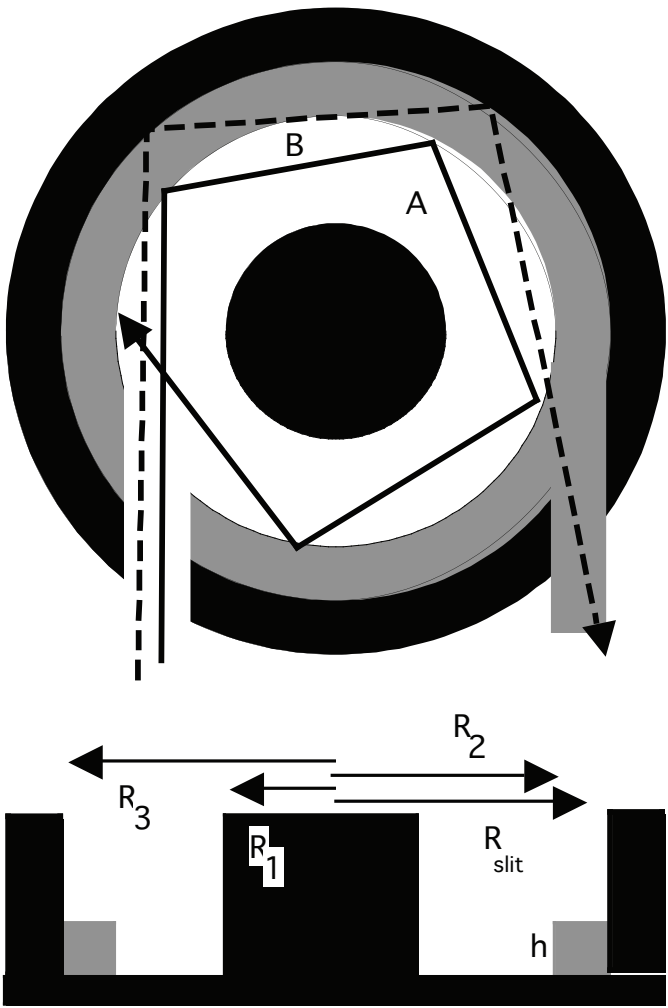}
\caption{\label{fig:fig1}The neutron centrifuge. A shows a classically confined orbit, B shows a higher energy one that escapes.}
\end{figure}

 	The basic concept is shown in Fig ~\ref{fig:fig1}
UCN's are injected into the centre of the apparatus. The height of the step h is adjusted so that the lowest energy bounce states will be trapped in the middle of the apparatus while the higher energy ones will bounce over the step and will be centrifuged out. It is obvious that this will work classically: it is less obvious that it will work quantum-mechanically.
 	In order to investigate this, we need to solve the 3-D Schrodinger equation which describes the neutron:
\begin{equation}-{{\hbar ^2} \over {2m}}\nabla ^2\Psi \left( {\vec r} \right)+\left( {\sigma z-E} \right)\Psi \left( {\vec r} \right)=0\  ,\  \sigma =mg\end{equation}

	In the case where the step-height h = 0, the wave function 
$\Psi _m(r,z,\phi )$
 can be separated to give 
\begin{equation}\Psi _{k\ln }(r,z,\phi )=f_{lk}\left( r \right)Z_n\left( z \right)e^{il\varphi }\end{equation}

	We now want to use these as the basis functions for the problem with a step. The step is defined in general via a function $H(r)$: in this case
\begin{eqnarray}\  H\left( r \right)_{}=0\  r<R_1\
&&\  H\left( r \right)_{}=h\  r\ge R_1\end{eqnarray}

We can then define the new wave  function as 
\begin{equation}\Phi _m(r,z,\phi )=\sum\limits_{}^{} {a_{k\ln }^m(r,z)}\Psi _{k\ln }(r,z,\phi )\end{equation}
The solution of this problem is shown in the appendix.

	To simulate the passage of a large number of neutrons through the centrifuge, we first choose a random superposition of the $\Phi _m(r,z)$. 
\begin{equation}{\Theta _0(r,z)=\sum\limits_{}^{} {b_m^0\Phi _m(r,z)}}\end{equation}
After passing once through the centrifuge, all components of the wave function lying beyond 
$r=R_s$ are eliminated, so we find a new wave function
\begin{equation}{\Theta _i(r,z)=\sum\limits_{}^{} {b_m^i\Phi _m(r,z)}}\end{equation}
where
\begin{equation}b_m^i=\sum\limits_{}^{} {d_{mm'}b_{m'}^{i-1}}\end{equation}
and
\begin{equation}d_{mm'}=\int_{}^{} {\Phi _m(r,z)\Phi _{m'}(r,z)}\theta \left( {R_s-r} \right)rdrdz\end{equation}

Note that the 
 $d_{mm'}$ only depend on the eigenfunctions, so it acts as a transfer matrix. The phases of the 
$b_m^i$ are randomized for each pass round the centrifuge. Strictly speaking, this is not adequate: instead we should form wave-packets and project out a superposition of the radial and bounce states. However, the process we have defined here is essentially equivalent. 

To obtain actual results, we must choose a set of parameters. We use $R_1 = 10 \mu, R_2 = 20 \mu, R_3 = 35 \mu, R_{slit} = 30 \mu$ and $h = 17 \mu$: these are fairly arbitrary, with the exception of h which must lie between the bounce heights for the ground and first excited states. The set of basis functions includes 12 bounce states and 6 radial states, for a total of 72: again this is not a critical parameter.

\begin{figure}
\includegraphics{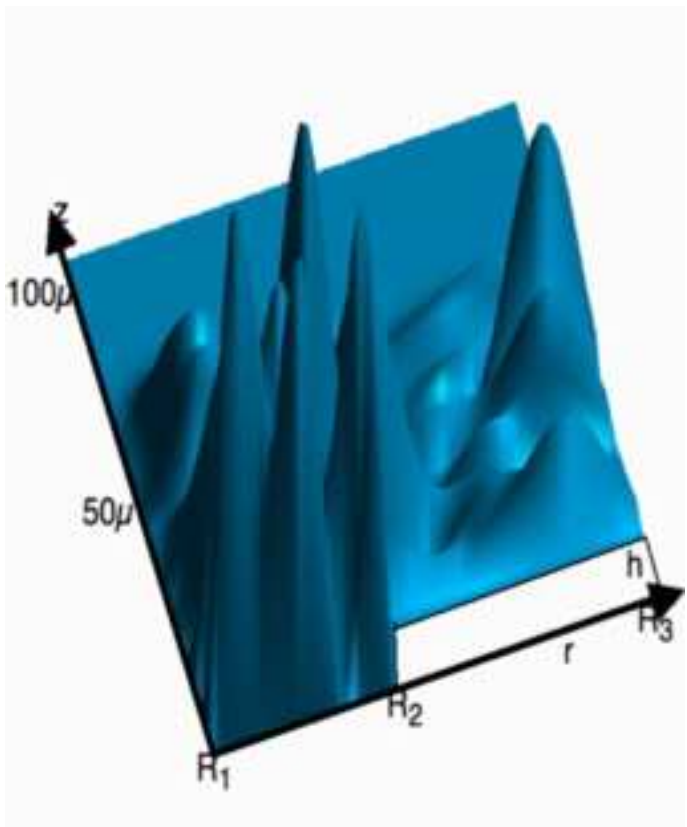}
\caption{\label{fig:fig2} initial probability distribution}
\end{figure}

\begin{figure}
\includegraphics{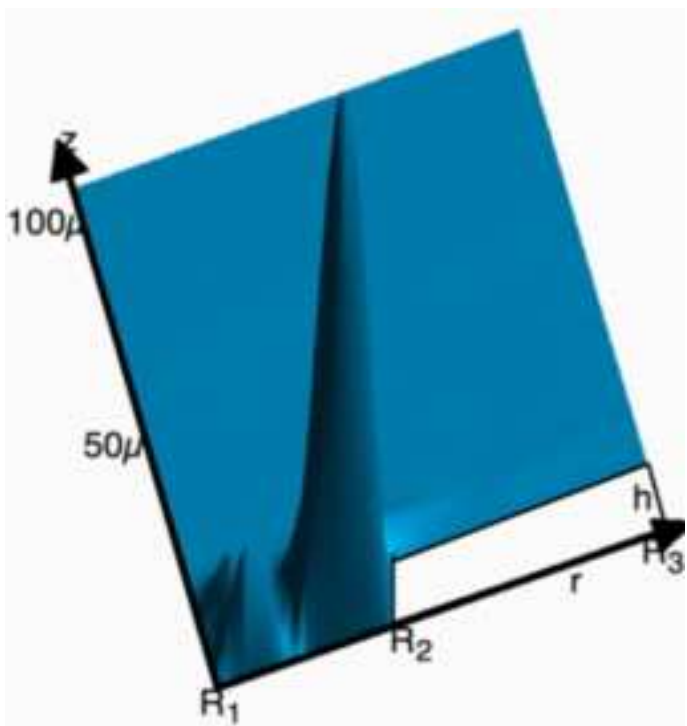}
\caption{\label{fig:fig3} Probability distribution after 100 passes}
\end{figure}

\begin{figure}
\includegraphics{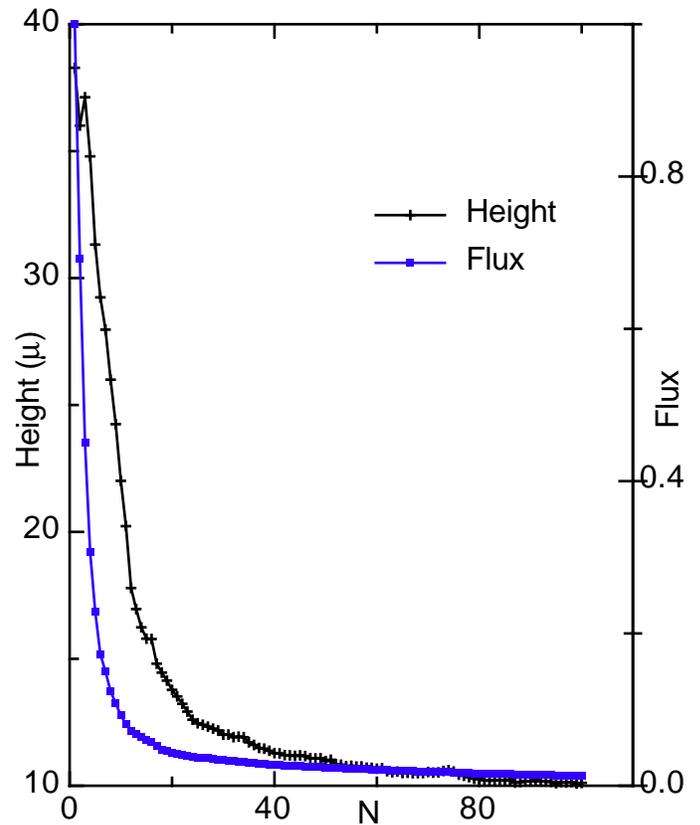}
\caption{\label{fig:fig4} Mean Height of state as a function of the number of bounces}
\end{figure}

The process is very efficient: in fig ~\ref{fig:fig2} 
 we show the initial probability  
$P_0(r,z)=\left| {\Theta _0(r,z)} \right|^2$ In in fig ~\ref{fig:fig3} we show the corresponding probability 
$P_{100}(r,z)$
after 100 passes.\ 
Obviously the wave function is very strongly concentrated in the central region, and corresponds on the lowest bounce states (n = 1). It is not straightforward to quantify this further, since our actual wavefunctions are a mixture of various bounce and radial states. However, it is possible to find the mean height after each passage through the apparatus, and compare this to the mean height of the lowest bounce state, which turns out to be 9.6 $\mu$. This is shown in fig ~\ref{fig:fig4}  where we plot the mean height as a function of the number of passes.
 These results are for l = 1: they are in fact more dramatic for larger l, as intuition would suggest.
	
	We conclude that this entirely passive device will perform a very effective job of separating the lowest bounce state. Two points should be noted, however. In practice we will not know if there is a neutron in the device. Secondly the step effectively mixes bounce and radial states and their energies, so that (e.g) each bounce state would be spread over a number of radial states. Careful adjustment of the radial parameters can minimize this.

\section {Finite penetration effects}
We have assumed above that the neutron is perfectly reflected. In fact this is not correct: the potential barrier at the metal has a finite height, given by the Fermi quasi-potential \cite{Golub}), which allows the wave-function to penetrate into the surface\footnote { The importance of this consideration was pointed out to me by Mike Pendlebury.}. This has two effects: firstly the calculated energies and bounce heights will not be correct. Secondly the penetration of the neutron allows it to be either absorbed or scattered inelastically by phonons: obviously either of these would destroy the state. Hence even the ground state would be expected to have a finite lifetime: if the lifetime is shorter than the neutron lifetime, it would seriously affect any potential experiments, so it is important to estimate it.

Provided that the penetration depth is small, we can estimate these effects by perturbation theory. The potential inside the metal is
\begin{equation}\delta V=\delta V_R+i\delta V_i\end{equation}
where 
\begin{equation}\delta V_R=252neV,\delta V_i=1.26peV\end{equation}
are the values appropriate for Be \cite {Golub}: other choices of material do not significantly affect the argument.
  The wave function for the n'th excited state inside the surface is given by
\begin{equation}\hat Z_n\left( z \right)=A_ne^{\kappa _nz}\left( {z<0} \right)\end{equation}

with 
\begin{equation}\kappa _n={1 \over \hbar }\sqrt {2m\left( {E_b-E_n} \right)}\end{equation}
where $E_b$ is the barrier height and $E_n$ is the energy of the n'th level.   Above the surface the wave function is given by
\begin{equation}\hat Z_n\left( z \right)=Z_n\left( {z+\delta z} \right)\left( {z>0} \right)\end{equation}
so that effectively the wave function has been shifted downwards by a small penetration depth $\delta z$.

 Obviously $\hat Z\left( z \right)$ must be continuous and differentiable at $z=0$, giving 
\begin{eqnarray}\delta z_n={1 \over {\kappa _n}} && A_n=\delta z_nZ'_n\left( 0 \right)=\delta z_n\beta Ai'\left( {-y_n} \right)\end{eqnarray}

This gives the probability of the neutron being inside the metal 
\begin{equation}P_n=\int\limits_{-\infty }^0 {\left| {\hat Z_n\left( z \right)} \right|^2dz={{A_n^2} \over {2\kappa _n}}}\end{equation}

Provided $P_n$ is small, the calculation is self-consistent. We find the penetration depth and the probability 
\begin{equation}\delta z_n\sim 0.01\mu ,P_n\sim 2\times 10^{-9}\end{equation}

Both increase very slowly with n, as classical intuition would suggest. This gives a correction to the real part of the energy 
\begin{equation}\delta E_n=P_nE_b\sim 0.01peV\end{equation}
which is less than a $1\% $
 change.
Potentially more important is the imaginary part of the potential, which gives rise to a lifetime 
\begin{equation}
\tau _n={\hbar  \over {2\int\limits_{-\infty }^0 {\left| {\hat Z_n\left( z \right)} \right|^2\delta V_idz}}}\sim 1.4\times 10^5s
\end{equation}

This is almost certainly over-optimistic, since the lifetime of higher energy UCN's is rather less than would be predicted by this method \cite{Byrne}. However, it does suggest that the confinement time for these states is likely to be at least comparable to the neutron lifetime of ~885 s.
 \section{Magnetic dipole interaction and the EDM measurement}

To utilize this apparatus, we assume that neutrons are trapped in this lowest state with all the higher energy states centrifuged out. 
The first observation is that it would be very easy to produce polarized neutron states by applying a static magnetic field. The magnetic and gravitational forces on the neutron are equal for $d_{mag}{{dB} \over {dz}}=mg$ or a modest ${{dB} \over {dz}}=1.7 Tm^{-1}$. Hence by applying such a field, one spin state would have its binding energy reduced, its bounce height increased and hence  would be centrifuged out , while the other spin state would be more strongly bound. Such a field could easily be achieved by a current loop surrounding the apparatus, although the spatial variation slightly complicates matters.

To study the state in more detail, we consider a technique of resonantly exciting a neutron in the n= 1 state to the n = 2 or higher. This could be done via a physical oscillation of the form discussed in \cite{Felber}, or an oscillating magnetic field. 
\begin{equation}\vec B(z,r)\ =\vec B_0\left( {z,r} \right)\cos \left( {\omega t} \right)\end{equation}

If the field is spatially uniform, the excitation matrix elements will vanish identically. We therefore assume that the magnetic field is produced by a loop of wire as above.
This gives rise to an interaction matrix element  of the form 
\begin{equation}T_n=\left\langle n \right|\vec \mu .\vec B\left| 1 \right\rangle \end{equation}

The probability for resonant transition to the first excited state (we follow \cite {Flugge} is given by
\begin{equation}P^{}={{\Omega ^2} \over {\gamma ^2}}\sin ^2\left( {\gamma t} \right)\end{equation}

where
\begin{equation}\gamma ^2=\Omega ^2+\delta \omega ^2,\Omega ^2={{T_n^2} \over {\hbar ^2}},\delta \omega ^2=\left( {\omega _1-\omega _2} \right)^2-\omega ^2\end{equation}
The frequency $\gamma $ must satisfy
\begin{equation}\gamma <<\left| {\omega _1-\omega _2} \right|=254Hz\end{equation}
which implies a maximum magnetic field of a few milligauss. Since the excited states will have an extremely small natural width, one could, in principle, measure the transition energy quite accurately.

To make an EDM measurement requires a field as large as possible, since the dipole moment is very small: we assume $E_{\max }=10^7Vm^{-1}$. To show what limits one could expect to put on the EDM, we suppose that the oscillating field $\vec E(z,r)\ =\vec E_0\left( {z,r} \right)\cos \left( {\omega t} \right)$ is supplied by a small charged sphere on the axis of symmetry of the apparatus, at a depth ${z_0}$. This gives rise to a perturbing potential of the form
\begin{equation}\vec d_{el}.\vec E={{d_{el}E_{\max }\left( {z+z_0} \right)\left( {R_1^2+z_0^2} \right)} \over {\left( {\left( {r+R_1} \right)^2+\left( {z+z_0} \right)^2} \right)^{3/2}}}\end{equation}

\begin{figure}
\includegraphics{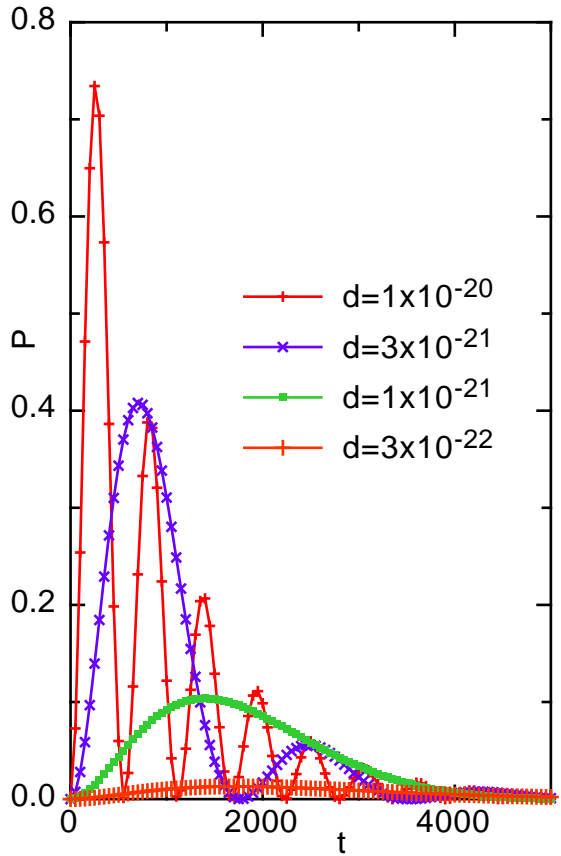}
\caption{\label{fig:fig5}Probability of excitation of the neutron from the n = 1 state into the n = 2 state for different values of the EDM}
\end{figure}

If $\delta\omega$ is set equal to zero, we obtain the transition probabilities shown in Fig ~\ref{fig:fig5}, for 4 values of the dipole moment $d_{el}$, where we have included the neutron lifetime. The transition time $1/\gamma$ for $d_{el}=3 \times 10^{-21}$ is approximately 1300 s, which is comparable to the neutron half-life. Clearly an optimized design would improve the result, but probably not by more than a factor of 10 at best.  Hence this technique cannot give a result which is competitive with the best current methods.

\section{Short distance behaviour of gravity}

Several models have suggested that the short distance behaviour of gravity may be modified from Newton's law. It has been pointed out by Bertolami and Nunes \cite{Bert} that, in principle, the system provides a method of measuring corrections to the gravitational interaction with a range of a few microns. We consider whether this could be done in practice.
The gravitational potential is taken to be

\begin{equation}V(r)={{Gm_1m_2} \over r}+{{Km_1m_2} \over r}e^{-\lambda r}\end{equation}

 A slab of material of density $\rho$ under the apparatus gives rise to a extra interaction the form
\begin{equation}\delta V(z)={{2\pi K\rho m_N} \over {\lambda ^2}}\left( {\lambda z+1} \right)e^{-\lambda z}\end{equation}
 This, of course, depends on our assumption of the slab being infinite: in practice, if the thickness of the slab is t, the approximation breaks down for $\lambda \sim t^{-1}$. This modifies the energy spectrum and hence in principle could be measured. For $\lambda <<h^{-1}$ (i.e. $\lambda \sim 10^4\mu ^{-1}$ or less), this term simplifies to give a constant term (which can be ignored, since it shifts all the levels equally) and a term of the form
\begin{equation} \delta V(z)=-\pi K\rho m_Nz^2 \end{equation}

which is independent of $\lambda $.
\begin{figure}
\includegraphics{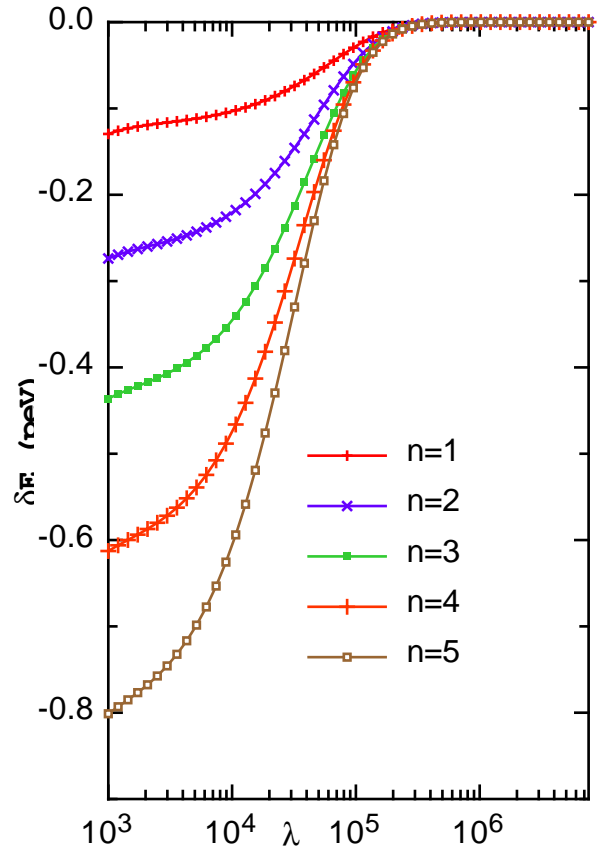}
\caption{\label{fig:fig6}Effect of extra gravitational interaction for  n=2, 3 and 4, as a function of the range  $\lambda$, with K = 1}
\end{figure}

In Fig ~\ref{fig:fig6} , we show the shift in the transition energy for the 5 lowest levels as a function of $\lambda $, for K = 1. We conclude that if the extra coupling constant was large and energy shifts could be measured down to 0.1 peV, it would be possible to look for an extra gravitational shift with a range of $\lambda ^{-1}\sim 1\mu $. 

One can again use resonant excitation of the ground state to put a better limit on the extra interaction. We assume that the slab is vibrated vertically with a frequency $\omega $ with an amplitude $z_0$ so that the distance from the top of the slab to the neutron is given by 
\begin{equation}\hat z\left( t \right)=z+z_0\left( {1-\sin \left( {\omega t} \right)} \right)\end{equation}
This means that the neutron would feel a time-varying potential of the form
\begin{equation}\delta V(z,t)={{2\pi K\rho m_N} \over {\lambda ^2}}\left( {\lambda \hat z\left( t \right)+1} \right)e^{-\lambda \hat z\left( t \right)}\end{equation}
which is obviously highly anharmonic. We can use the expansions \cite{Abram1}
\begin{equation}e^{\lambda z_0\cos \left( {\omega t} \right)}=I_o\left( {\lambda z_0} \right)+2\sum\limits_{k=0}^\infty  {I_k\left( {\lambda z_0} \right)}\cos \left( {\omega t} \right)\end{equation}

Since the transition energy to the n'th excited bounce state $\omega _1-\omega _n$ is not a simple multiple of $\omega =\left| {\omega _1-\omega _2} \right|$
, the resonance condition means that we can ignore all terms except one in the expansion.  This gives rise to an interaction of the form
\begin{eqnarray}\delta V(z,t)=&&\sin \left( {\omega t} \right){{2\pi K\rho m_N} \over {\lambda ^2}}e^{-\lambda \left( {z+z_0} \right)}\\
&&\times \left( {\left( {\lambda \left( {2z+z_0} \right)+4} \right)I_1\left( {\lambda z_0} \right)-2\lambda z_0I_1\left( {\lambda z_0} \right)} \right) \nonumber
\end{eqnarray}

This gives rise to an transition time which depends linearly on K and non-linearly on $\lambda$. 
\begin{figure}
\includegraphics{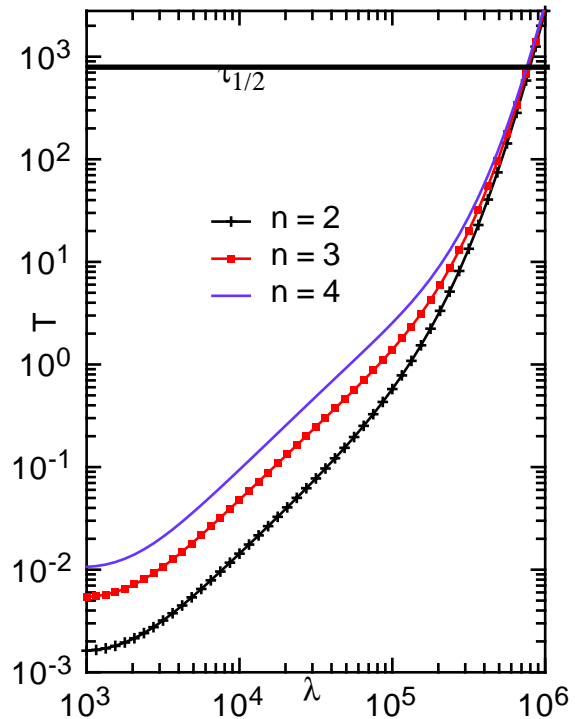}
\caption{\label{fig:fig7}Plot of the transition time t as a function of $\lambda$ from the ground state to n=2,3 and 4. The neutron half-life is also shown.}
\end{figure}

In Fig ~\ref{fig:fig7} we show this time as a function of $\lambda$ for K=1. It would be possible to use this to investigate short range changes to the gravitational interaction, but this does not allow a measurement to be extended down into the the phenomologically interesting range of $K\approx G$.

 \section{Conclusions}	
 We conclude that it is possible to isolate this novel quantum state, which, of course, has its own intrinsic interest, since the number of simple quantum mechanical states is small, and this, along with Rydberg states, are unique in having a macroscopic size. It is obviously disappointing that it does not seem possible to set new limits on the neutron EDM, but it does appear possible to test the gravitational interaction at very short distances. 

 We do  not underestimate the problems of performing a real experimental measurement. It would obviously be very hard to observe the bouncing neutron at all, given the low flux rates combined with the huge flux of more energetic neutrons. Even if a neutron is captured in the n=1 state, it will then be quite difficult to excite it and detect it on its escape from the centrifuge.

There is no reason in principle why this method cannot be extended to atoms. Atomic fountain experiments provide atoms which can have temperatures as low as 1 nK, and hence would provided a much more numerous source. The major unknown is that atom-surface Van der Waals interactions are likely to be fairly strong and attractive. This would seriously distort the lowest bounce states. However, atoms are stable and cannot easily be absorbed.

Finally we note that, with the addition of an appropriate static magnetic field, it would be possible to produce quantum states with an even lower energy in a controlled fashion. 
\begin{acknowledgments}

I am very grateful to Bill Romo and Mike Pendlebury for their comments, and particularly to M. K. Sundaresan for pointing out an error in an earlier version of this work and advice on the solution given in the appendix. I would like to thank Frank Close and his colleagues at Oxford for their hospitality. Financial support from NSERC is gratefully acknowledged.
\end{acknowledgments}

\appendix
\section {Numerical solution of the Schrodinger equation }
The unperturbed wave function can be written
\begin{equation}\Psi _{k\ln }(r,z,\phi )=f_{lk}\left( r \right)Z_n\left( z \right)e^{il\varphi }\end{equation}

The radial function has to vanish at 
$r=R_1$ and $r=R_3$. The first gives the solution 
\begin{equation}f(r)=Y_l\left( {\kappa R_1} \right)J_l\left( {\kappa r} \right)-J_l\left( {\kappa R_1} \right)Y_l\left( {\kappa r} \right)\end{equation}

and satisfying the second condition defines the eigenvalues 
$\kappa _{kl}$
 via 
\begin{equation}Y_l\left( {\kappa _{kl}R_1} \right)J_l\left( {\kappa _{kl}R_3} \right)-J_l\left( {\kappa _{kl}R_1} \right)Y_l\left( {\kappa _{kl}R_3} \right)=0\end{equation}

 This is solved numerically: it is easy to get an asymptotic form for the eigenvalues: 
\begin{equation}\kappa _{kl}={{k\pi } \over {R_3-R_1}}\end{equation}

We must now find the perturbed wave functions $\Phi _{m}(r,z)$ for the problem with a step in terms of the unperturbed wave-functions $\Psi _{k\ln }(r,z)$ (the dependence on $\phi $ is irrelevant). To satisfy the boundary conditions with the step, we write
\begin{eqnarray}
\Phi _m(r,z)=  \sum\limits_{}^{} {a_{k\ln }^m} \\
\left( {\psi _{k\ln }\left( {r,z} \right)-t\left( {r,z,z_s\left( r \right)} \right)\psi _{k\ln }\left( {r,z_s\left( r \right)} \right)} \right) \nonumber \\
\end{eqnarray}
or
\begin{equation}
\Phi _m(r,z)=\sum\limits_{}^{} {a_{k\ln }^m}\hat \psi _{k\ln }\left( {r,z,z_s\left( r \right)} \right)
\end{equation}

The function 
${t\left( {r,z,z_s\left( r \right)} \right)}$
 must satisfy 
$t\left( {r,z_s\left( r \right),z_s\left( r \right)} \right)=1$.
so that 
$\hat \psi _{k,n}\left( {r,z_s\left( r \right),z_s\left( r \right)} \right)=0$,
must vanish as 
$z\to \infty $ at least as fast as the Airy function itself, and must vanish as $r\to R_1 $. It must also be smooth. Apart from this it is quite arbitrary: we use 
\begin{equation}
t\left( {r,z,z_s\left( r \right)} \right)
\begin{array}{c}{={{(R_2-r)} \over {(R_2-R_1)}}\   z<h,r<R_2} \\
{={{(u_h-u)} \over {u_h}}e^{-\beta ^2\left( {z-h)^2} \right)}\   z\ge h,r<R_1}\\
{=e^{-\beta ^2\left( {z-h)^2} \right)}z\ge h,r\ge R_1}
\end{array};,
\end{equation}

where 
\begin{eqnarray}
u=\left( {\left( {z-h} \right)^2+\left( {R_2-r} \right)^2} \right)^{1/2} \nonumber\\
u_h=\left( {R_2-R_1} \right)\left( {\left( {{{z-h} \over {R_2-r}}} \right)^2+1} \right)^{1/2}
\end{eqnarray}

which satisfies these properties.

These states have energy given by 
\begin{equation}\hat E_m={{\left\langle {\Phi _m^{}|-{{\hbar ^2} \over {2m}}\nabla ^2+\sigma z|\Phi _m^{}} \right\rangle } \over {\left\langle {\Phi _m^{}|\Phi _m^{}} \right\rangle }}\end{equation}
 or

\begin{eqnarray}\hat E_m\sum\limits_{kl} {\sum\limits_{k'l'} {a_{kl}^ma_{kl}^m\int {\hat \psi _{kl}(\vec r)\hat \psi _{kl}(\vec r)d\vec r}}} \nonumber\\
\end{eqnarray}

The principal problem is caused by evaluating
$$\nabla ^2\hat \psi _{k,n}\left( {r,z,z_s\left( r \right)} \right)$$. This must in general be done numerically. 
	We regard the 
$a_k^{}$
as variational parameters and use them to minimize $\hat E_m$. This gives 
\begin{equation}\hat E_m\sum\limits_0^N {a_k^{}}G_{kl}={1 \over 2}\sum\limits_0^N {a_k^{}}\left( {H_{kl}+H_{lk}} \right)\end{equation}
where 
\begin{equation}G_{kl}=\int\limits_{}^{} {\hat \psi _k\hat \psi _l}drdz,\  H_{kl}=\int\limits_{}^{} {\hat \psi _kH\hat \psi _l(x)}drdz\end{equation}
which is an eigenvalue equation for the $\hat E_m$. Note that $G_{kl}$ is symmetric but $H_{kl}$ is not. However, the resulting eigenvalue equation contains only symmetric matrices and can be solved by the techniques of \cite {NumRec}.
\end{document}